\documentclass[12pt]{article}\textheight 24.5cm \textwidth 17cm\voffset=-1.in\hoffset= - 1.in
\makeatletter
\@addtoreset{equation}{section}

\makeatother
\usepackage[utf8]{inputenc}
\usepackage[english]{babel}
\usepackage[T2A]{fontenc}
\usepackage{amsmath}
\usepackage{slashed}
\usepackage{graphicx}
\usepackage{amsfonts}
\usepackage{amssymb}
\usepackage{psfrag}
\usepackage{hhline}
\usepackage{cite}
\usepackage{upgreek}
\usepackage{epsfig,amsfonts}

\newdimen\tableauside\tableauside=.5ex
\newdimen\tableaurule\tableaurule=0.4pt
\newdimen\tableaustep
\def\phantomhrule#1{\hbox{\vbox to0pt{\hrule height\tableaurule
width#1\vss}}}
\def\phantomvrule#1{\vbox{\hbox to0pt{\vrule width\tableaurule
height#1\hss}}}
\def\sqr{\vbox{%
  \phantomhrule\tableaustep

\hbox{\phantomvrule\tableaustep\kern\tableaustep\phantomvrule\tableaustep}%
  \hbox{\vbox{\phantomhrule\tableauside}\kern-\tableaurule}}}
\def\squares#1{\hbox{\count0=#1\noindent\loop\sqr
  \advance\count0 by-1 \ifnum\count0>0\repeat}}
\def\tableau#1{\vcenter{\offinterlineskip
  \tableaustep=\tableauside\advance\tableaustep by-\tableaurule
  \kern\normallineskip\hbox
    {\kern\normallineskip\vbox
      {\gettableau#1 0 }%
     \kern\normallineskip\kern\tableaurule}%
  \kern\normallineskip\kern\tableaurule}}
\def\gettableau#1 {\ifnum#1=0\let\next=\null\else
  \squares{#1}\let\next=\gettableau\fi\next}

\author{ A. Belavin\footnote{ belavin@itp.ac.ru}$\,\,^{1,2,3}$, L. Spodyneiko\footnote{lionspo@itp.ac.ru}$\,\,^{1}$
\vspace*{10pt}\\[\medskipamount]
$^1$~\parbox[t]{0.88\textwidth}{\normalsize\it\raggedright
L.D.Landau Institute for Theoretical Physics,
142432 Chernogolovka, Russia}
\vspace*{10pt}\\[\medskipamount]
$^2$~\parbox[t]{0.88\textwidth}{\normalsize\it\raggedright
Moscow Institute of Physics and Technology, 141700 Dolgoprudny, Russia}
\vspace*{10pt}\\[\medskipamount]
$^3$~\parbox[t]{0.88\textwidth}{\normalsize\it\raggedright
Institute for Information Transmission Problems, 127994 Moscow, Russia}
}
\date{}
\title{\bf Flat structures on Frobenius Manifolds in the case of  irrelevant deformations}
\begin{document}
\maketitle
\begin{abstract}
In this paper we use the  recently suggested  conjecture about the integral representation for  the flat coordinates on  Frobenius manifolds, connected with the isolated singularities,   to compute the flat coordinates and Saito primitive form on the space of the deformations of Gepner chiral ring $\widehat{SU}(3)_4$.

 We verify this conjecture  comparing the expressions for the flat coordinates obtained from the conjecture with the one found by direct computation. The considered case is of a particular interest since together with the relevant and marginal deformations it has one irrelevant deformation.

\end{abstract}
\vspace*{10pt}

\section{Introduction}
This paper is a continuation of the previous series of works \cite{BGK,BB1,BB2} and dedicated to a new approach to computations of the flat coordinates of the Frobenius manifolds connected
with isolated singularity \cite{Dub, Saito}.

Frobenius structure arises in the  three kinds of models of QFT and string theory. Namely, in the models of  two-dimensional topological CFT \cite{DVV}, in the models of space-time supersymmetric  compactifications of string theory on Calabi-Yau manifolds \cite{CHSW,Gep,COGP,BCOFHJQ} and  in the models of Polyakov noncritical string theory \cite{BDM}.

One of the key ingredients for the solution of this models is knowledge \cite{DVV, VB,VB1} of the flat  coordinates on the corresponding Frobenius manifolds.

A new method for the computation of the flat coordinates, based on the conjecture about an integral representation, has been suggested for the models connected with  the simple ADE singularities  in \cite{BGK}. In \cite{BB1,BB2} this method and the conjecture itself were formulated for the case  of the general isolated singularity. Also, in these works the conjecture was verified for  the model with a number of relevant and one marginal deformations.

The aim of this work is to use  and verify the conjecture in  the computation of  the flat coordinates on Frobenius manifolds of  deformations the Gepner chiral  ring $\widehat{SU}(3)_4$.
This model has one marginal and one irrelevant deformation. We verify the results obtained by the use of the conjecture by comparing with  the results of the  direct computation.

In section 2 we briefly review Dubrovin-Saito theory. In section 3 we formulate the conjecture of the integral representation for flat coordinates  in case when Saito primitive form is not trivial what happens when the marginal and irrelevant deformations take place. In section 4 we review some needed  facts about the deformation of the Gepner chiral ring  $\widehat{SU}(3)_4$  for which our computations are performed. In section 5 we explain how one can find metrics and flat coordinates by a direct way from the Frobenius manifold structure axioms, compare the results of both computations and find  their coincidence. In section 6 we discuss moduli of the primitive form and their connection with the resonances in the ring. We provide more detailed expressions for the flat coordinates and the primitive form in the appendix.

\section{Preliminaries}\label{sec Prelim}
In this section, we review  the role of the flat coordinates of the Frobenius manifold for the case of topological CFT which comes from Witten twist and restriction of space of states on the chiral sector of the $N=2$ SCTF Landau-Ginzburg model.

  In these models superpotential $W_0[\Phi_1,\dots,\Phi_n]$ depends on $n$ fundamental chiral fields. These fields generated chiral ring $R_0$ and we will denote basis in it as $\Phi_\alpha$ for $\alpha=1,\dots,M.$ Here $M=\dim R_0$, first fields with $\alpha=1,\dots,n$ will label generators of the ring and $\Phi_1=1$ is a unity operator.

  The chiral ring $R_0$ is isomorphic  to the ring of the polynomials of $x_i$
  \begin{equation}\label{chiral ring}
  R_0=\mathbb C^n[x_1,\dots,x_n]\Big/\left\{\frac{\partial W_0}{\partial x_i}\right\},
  \end{equation}
where $\left\{\frac{\partial W_0}{\partial x_i}\right\}$ denotes the ideal generated by the partial derivatives of the polynomial $W_0[x_i]$.

It was shown in \cite{DVV} that for computation of the correlation functions of the fields $\Phi_\alpha$ and its superpartners $\Phi_\alpha^{(1,1)}=G_{-1/2}^-G_{-1/2}^+\Phi_\alpha$ it is needed and sufficient to know the two-point functions
\begin{equation}
\eta_{\alpha\beta}=\langle \Phi_\alpha \Phi_\beta \rangle,
\end{equation}
together with the perturbed three-point function
\begin{equation}
C_{\alpha\beta\gamma}(s_1,\dots,s_M)\stackrel{\text{def}}{=}\langle \Phi_\alpha\Phi_\beta\Phi_\gamma\exp\left( \sum_{\lambda=1}^M s_\lambda \int \Phi^{(1,1)}_\lambda\,d^2z \right)\rangle.
\end{equation}

It was also shown in \cite{DVV} that $\eta_{\alpha \beta}$ is non-degenerate and $s$-independent and  $C_{\alpha\beta\gamma}(s)$ can be expressed through a prepotential (or free eneregy) $\mathcal F$
\begin{equation}
C_{\alpha\beta\gamma}=\frac{\partial^3\mathcal F}{\partial s_\alpha s_\beta s_\gamma}.
\end{equation}
At last $C^{\gamma}_{\alpha\beta}\stackrel{\text{def}}{=}\eta^{\gamma\delta}C_{\alpha\beta\delta}$ are subject of the equation
\begin{equation}
C^\rho_{\alpha\beta}C^\mu_{\rho\gamma}=C^\rho_{\alpha\gamma}C^\mu_{\rho\beta}.
\end{equation}
At $s_\alpha=0$ this ring coincide with the chiral ring $R_0$ defined by (\ref{chiral ring}).

These relations together with the evident property $C^{\gamma}_{\alpha\beta}=C^{\gamma}_{\beta\alpha}$ mean that $C^\gamma_{\alpha\beta}(s)$ are structure constants for a commutative, associative algebra or a ring $R$ with unity  which depends on the parameters $\{s_\alpha\}$.

The properties of $\eta_{\alpha\beta}$ and $C^\gamma_{\alpha\beta}(s)$ mean indeed that we have the Frobenius manifold structure \cite{Dub} and $s_\mu$ are nothing but the flat coordinates on this manifold, i.e. such coordinates in which the Riemann metric $\eta_{\alpha\beta}$ is constant.

The crucial fact, found in \cite{DVV},
 which makes possible to exactly solve the topological models of such kind, is that the Frobenius manifold, defined by $C_{\alpha\beta\gamma}(s)$ and $\eta_{\alpha\beta}$, coincides with a Frobenis manifold  defined by the versal deformations $W(x,t)$ of a  superpotential $W_0$
\begin{equation}
W(x,t)\stackrel{\text{def}}{=}W_0(x)+\sum_{\alpha=1}^M t_\alpha e_\alpha(x).
\end{equation}

Here $\{e_\alpha\}$ is a basis of the ring $R_0$ (\ref{chiral ring}), $e_1(x)=1$ is a unity element of $R_0$.
The corresponding ring, defined by $W(x,t)$ as the ring of polynomials of $x_i$,
\begin{equation}
R_W=\mathbb C^n[x_1,\dots,x_n]\Big/\left\{\frac{\partial W}{\partial x_i}\right\}.
\end{equation}
The structure constants $\widetilde C^\gamma_{\alpha\beta}(t)$ of $R_W$ in the basis $\{e_\alpha\}$ are defined by the relations
\begin{equation}
e_\alpha e_\beta=\widetilde C^\gamma_{\alpha\beta}(t)e_\gamma \mod \{\frac{\partial W}{\partial x_i}\}.
\end{equation}
The Riemann metric $g_{\alpha \beta}(t)$ is defined as the Grotendick residue in the terms of the Saito primitive form \cite{Saito}

\begin{equation}
\Omega(x,t)=\lambda(x,t)dx_1\wedge \dots\wedge x_n
\end{equation}
as follows
\begin{equation}
g_{\mu\nu}= \text{Res}_{x=\infty} \frac {e_\mu e_\nu\Omega}{\prod_i \partial W/\partial x_i}.
\end{equation}
It was proved in \cite{M.Saito} the  primitive form does exist. Namely, there exist such a differential form $\Omega(x,t)$, that the structure constants $\widetilde C^\gamma_{\alpha\beta}(t)$ (2.8) and Riemann metric $g_{\alpha \beta}(t)$ (2.10) satisfy the Dubrovin Frobenius manifold axioms:
\begin{align}
\widetilde C^\rho_{\alpha\beta}\widetilde C^\mu_{\rho\gamma}&=\widetilde C^\rho_{\alpha\gamma}\widetilde C^\mu_{\rho\beta},\\
\label{curve}R_{\mu\nu\lambda\sigma}[g_{\alpha\beta}]&=0,\\
\label{der}\nabla_\sigma \widetilde C_{\mu\nu\lambda}&=\nabla_\mu \widetilde C_{\sigma\nu\lambda},\\
\label{sym}\widetilde C_{\mu\nu\lambda}&=\widetilde C_{\nu\mu\lambda}=\widetilde C_{\mu\lambda\nu}.
\end{align}
The deformation parameters $\{t_\alpha\}$ are some coordinates on the Frobenius manifold.
The coupling constants  $s^\mu$ are  the flat coordinates on it. They are functions of the deformation parameters~$\{t_\alpha\}$.

The knowledge of these functions permit to express the perturbed three-point functions
$C_{\alpha\beta\gamma}$  and the prepotential $\mathcal F$ in terms of $g_{\mu\nu}$
and  $\widetilde C^\gamma_{\alpha\beta}(t)$ .
Thus  the determination of the functions $s^\mu(t)$ and the primitive form $\Omega(x,t)$ is the major part of the solution of the topological Landau-Ginzburg model.

\section{The  flat coordinates through the oscillating integrals.}\label{sec conj}

We will assume that the superpotential $W_0(x)$ is a quasihomogeneous polynomial associated to an isolated singularity
\begin{equation}
W_0(\Lambda^{\rho_i} x_i )= \Lambda^d  W_0(x_i),
\end{equation}
where  integer weights  $d=[W_0]$ and $ \rho_i=[x_i] $.

In this case we can choose the basis $e_\alpha$ of the ring $R_W$ to be quasihomogeneous . We will denote its weights as ${\deg e_\alpha}$. The elements of the basis are called relevant, marginal or irrelevant if their weights satisfy correspondingly the relations ${\deg e_\alpha<d}$, ${\deg e_\alpha=d}$ or  ${\deg e_\alpha>d}$.

It was conjectured in \cite{BGK,BB2} that the flat coordinates are given by the following integral expression
\begin{equation}\label{conj}
s_\mu(t)=\sum_{m_\alpha \in \Sigma_\mu}\left(\int_{\gamma_\mu}\exp(W_0(x))\prod_\alpha e_\alpha^{m_\alpha} \Omega\right)\prod_\alpha \frac {t_\alpha^{m_\alpha}}{m_\alpha!},
\end{equation}
where $\Sigma_\mu$ is specified by requirement for l.h.s. and r.h.s. of this equation to have the same weights. We will give the explicit expression for it below.

The cycles $\gamma_\mu$ form basis
 for   the homology $ H_n({\mathbb C}^n, \operatorname{Re}W_{0}=-\infty)$ which defined

\noindent as~$\lim_{L\to +\infty} H_n ({\mathbb C}^n / \{ \operatorname{Re}W_0 \leq -L\}) $.\footnote{In this limit, a homology class can be repesented by non-compact closed $n$-dimensional submanifold $\gamma$ in ${\mathbb C}^n$ such that $\operatorname{Re}(W_0)$ tends to $-\infty$ at infinity, and therefore$\int_{\gamma} e^{W_0} dx$ converges.}

A simple  example of the explicit choice of such cycles  for $n=1$ has been  given in \cite{BGK} on
Figure 2. It  illustrates this notion.

We fix the normalization of the coordinates by the requirement for the first term of the decomposition to be $s_\mu=t_\mu+\dots$.

For computing the  integrals in (3.2) we  use the same way as in \cite{BB2}. The main point of the computation is the following property of the oscillating integrals

\begin{equation}
\int_\gamma \exp(W_0(x)) P_1(x)dx=\int_\gamma \exp(W_0(x)) P_2(x)dx,
\end{equation}
if there exist an $(n-1)$-form $U$ such that
\begin{equation}
P_1(x)dx=P_2(x)dx+D_{W_0}U,
\end{equation}
where $D_{W_0}$ is Saito differential
\begin{equation}
D_{W_0}=d+dW_0\wedge.
\end{equation}
The differential $D_{W_0}$ defines the Saito cohomology $ H^n$ on the space of n-forms.

The forms  $e_\mu dx$ for $\mu=1,\dots,M$ can be chosen as a convenient basis in $ H^n$.

Let us define a pairing between the elements of the basis $e_\mu dx$ in  $ H^n$   and
the cycles $\gamma_\mu$ as
\begin{equation}\label{hom bas}
 r_{\mu,\nu}=\int_{\gamma_\mu}\exp(W_0)e_\nu dx.
\end{equation}
We can choose the homology basis $\gamma_\mu$ to be dual to the cohomology basis  $ e_\mu dx $.

The simplest choice of the dual basis is tempting to be
\begin{equation}\label{can bas}
 r_{\mu,\nu}\stackrel{\text{?}}{=}\delta_{\mu,\nu}.
\end{equation}
However, a more general possibility have to be considered. The reason for this is  the occurrence of resonances. We will call resonance the case when the  weights of some coordinates satisfy $[s_\mu]-[s_\nu]= 0 \mod d$ and $[s_\mu]\ne[s_\nu]$.  We use the following choice
$r_{\mu,\mu}=1$ for all $\mu$ and $r_{\mu,\nu}=0$ if coordinates $s_\mu$ and $s_\nu$ are not in resonance.\footnote{In the case of our interest $\widehat{SU}(3)_4$ this gives four parameters $r_{1,14},r_{2,15},r_{14,1},r_{15,2}$.  The last two are fixed by  the normalization condition and the equation (\ref{t1 eq}). However, $r_{1,14},r_{2,15}$ stay to be free parameters.} Doing in this way we get the expressions for the flat coordinates which depend on some extra parameters as it is predicted in \cite{M.Saito1}.

From dimensional reasoning the primitive form $\Omega$ must be decomposed as
\begin{equation}\label{pr form}
\Omega=\sum_{n,l\in \omega}A(n,l) \prod_\alpha e^{n_\alpha}_\alpha t_\alpha^{l_\alpha}dx,
\end{equation}
where the summation domain $\omega$ is defined as
\begin{equation}
\omega: \quad \sum_{\alpha}( n_\alpha [e_\alpha]+l_\alpha [t_\alpha])=0, \quad n_\alpha\ge0,\quad l_\alpha\ge0.
\end{equation}
Substituting (\ref{pr form}) into (\ref{conj}) one finds
\begin{equation}\label{conj 2}
s_\mu(t)=\sum_{\substack{m_\alpha\in \Sigma_\mu,\\ n_\alpha,l_\alpha \in \omega}} \left(\int_{\gamma_\mu}\exp(W_0(x))A(n,l)\prod_\alpha e_\alpha^{m_\alpha+n_\alpha} dx\right)\prod_\alpha \frac {t_\alpha^{m_\alpha+l_\alpha}}{m_\alpha!}.
\end{equation}
Now, we can give the expression for $\Sigma_\mu$ more explicitly
\begin{equation}
\Sigma_\mu: \quad \sum_{\alpha} (m_\alpha+l_\alpha) [t_\alpha]=[s_\mu],
\quad m_\alpha\ge 0.
\end{equation}

Since $e_\mu dx$ form a basis of $ H^n$, any n-form can be decomposed in it. In particular,
\begin{equation}\label{B coef}
\prod_{\alpha}e_\alpha^{k_\alpha}dx = \sum_{\mu}B_{\mu}(k) e_\mu dx+D_{W_0}U.
\end{equation}
From the homogeneity requirements only such elements $ e_\mu dx$ of the basis appear in the r.h.s of this equation whose weights are equal to those of the l.h.s module $d$. In case of the resonance a few elements of the same weights  can appear in r.h.s. of (\ref{B coef}). Their appearence in the oscilating integrals (\ref{hom bas}) is the reason of arising of the parameters $ r_{\mu,\nu}$ in the expressions for $s_\mu$  when  $ e_\mu$ and  $ e_\nu$ are in a resonace.

In the case of our interest $\widehat{SU}(3)_4$, which is considered below, there are two resonances $[s_1]-[s_{14}]=7$ and $[s_2]-[s_{15}]=7$. We find that in this case  two  parameters $r_{1,14}$ and $r_{2,15}$, if they are not assumed to be equal zero,  arise in the expressions for the flat coordinates derived from  (3.2).

For given $k_\alpha$ we can solve  the equation (\ref{B coef}) and find the coefficients $ B_{\mu}(k)$ .
Substitution them into (\ref{conj 2}) gives
\begin{equation}\label{conj 3}
s_\mu(t)=\sum_{\substack{m_\alpha\in \Sigma_\mu,\\ n_\alpha,l_\alpha \in \omega}} A(n,l) B_\mu(m+n) \prod_\alpha \frac {t_\alpha^{m_\alpha+l_\alpha}}{m_\alpha!}.
\end{equation}
This formula gives expressions for $s_\mu$ that depend on the unknown  parameters $A(n,l)$ of the primitive form.
We  find these parameters  using  the normalisation conditions together with the equation  \cite{BB2}
\begin{align}\label{t1 eq}
\frac{\partial s_\mu}{\partial t_1}=\delta_{\mu,1}.
\end{align}
In such a way we arrive to the  explicit expression for flat coordinates.
 The final answer for the flat coordinates contains no free parameters besides  those of  $r_{\mu,\nu}$, which correspond to the resonances.

\section{The deformed  chiral ring $\widehat{SU}(3)_4$}
In the topological CFT,which is connected with the deformed chiral ring $\widehat{SU}(3)_4$ \cite{Gep1},
the superpotential is
\begin{equation}
W_0(x_1,x_2)=\frac{q_1^7+q_2^7}7,
\end{equation}
where $x_1=q_1+q_2$, $x_2=q_1q_2$.

We choose the basis of the ring to be Schur polynomials \cite{BB2}
\begin{align}
e_1\equiv e_{\varnothing}=1\;, \quad
e_2\equiv e_{\tableau{1}}=q_1+q_2\;, \quad
e_3\equiv e_{\tableau{1 1}}=q_1 q_2\;. \quad
e_4\equiv e_{\tableau{2}}=q_1^2+q_1 q_2+q_2^2\;, \quad \nonumber\\
e_5\equiv e_{\tableau{2 1}}=q_1 q_2(q_1+q_2)\;, \quad
e_6\equiv e_{\tableau{3}}=q_1^3+q_1^2 q_2+q_1 q_2^2+q_2^3\;,\quad
e_7\equiv e_{\tableau{2 2}}=q_1^2q_2^2\;,\quad\nonumber\\
e_8\equiv e_{\tableau{3 1}}=q_1 q_2(q_1^2+q_1 q_2+q_2^2)\;,\quad
e_9\equiv e_{\tableau{4}}=q_1^4+q_1^3q_2+q_1^2q_2^2+q_1q_2^3+q_2^4\;,\quad
\;\quad\nonumber\\
e_{10}\equiv e_{\tableau{3 2}}=q_1^2 q_2^2(q_1+q_2)\;,\quad
e_{11}\equiv e_{\tableau{4 1}}=q_1 q_2(q_1^3+q_1^2 q_2+q_1 q_2^2+q_2^3)\;,\quad
e_{12}\equiv e_{\tableau{3 3}}=q_1^3 q_2^3\;,\quad\nonumber\\
e_{13}\equiv e_{\tableau{4 2}}=q_1^2 q_2^2(q_1^2+q_1 q_2+q_2^2)\;,\quad
e_{14}\equiv e_{\tableau{4 3}}=q_1^3 q_2^3(q_1+q_2)\;,\quad
e_{15}\equiv e_{\tableau{4 4}}=q_1^4q_2^4. \nonumber
\end{align}
The deformed superpotential is
\begin{equation}
W(x_1,x_2)=\frac{q_1^7+q_2^7}7+\sum_{\mu=1}^{15}t_\mu e_\mu
\end{equation}
The first 13 elements of this basis are related to relevant deformations. The elements $e_{14}$  and  $e_{15}$  are related to the marginal and irrelevant deformations correspondingly.

We computed  the flat coordinates up to the 6th order   in  $t $  by  using the technique of the previous section.
The expressions for them  up to the 2nd order are presented in the appendix.
We also put there  the answer for the primitive form $\Omega$ up to the 2nd order in $t$.

\section{Direct computation of flat coordinates}\label{dir comp}
The expression for the flat coordinates (\ref{conj 3}) is a conjecture. This conjecture was tested in \cite{BB1,BB2} for  the topological CFT connected with  chiral  ring $\widehat{SU}(3)_3$, where one marginal deformation takes place.

One of the main aims of this work is to check this conjecture for the case when  there are also irrelevant deformations like it takes place for the model connected with the deformed Gepner chiral  ring $\widehat{SU}(3)_4$ .

In order to do it, we have to compute the flat coordinates by the direct way. We did it perturbatively in overall $t$ up to 4th order in $t$. The final answers are too lengthy to be presented here. Therefore, we will only outline the main steps of the calculation giving as many details as possible.

The metric on the Frobenius manifold is defined as
\begin{equation}
g_{\mu\nu}= \text{Res}_{x=\infty} \frac {e_\mu e_\nu\Omega}{\prod_i \partial W/\partial x_i}.
\end{equation}
Instead of computing  this residue we will follow the  way used in \cite{Klemm}. Namely we rewrite the metric as
\begin{equation}
g_{\mu\nu}=C^\lambda_{\mu\nu}(t)\text{Res}_{x=\infty} \frac {e_\lambda\Omega}{\prod_i \partial W/\partial x_i} = C^\lambda_{\mu\nu}(t) h_\lambda(t),
\end{equation}
where $h_\mu(t)$ are some unknown functions of $t$. These functions can be found from the Frobenius axioms
\begin{align}
\label{curve}R_{\mu\nu\lambda\sigma}[g_{\alpha\beta}]&=0,\\
\label{der}\nabla_\sigma C_{\mu\nu\lambda}&=\nabla_\mu C_{\sigma\nu\lambda},\\
\label{sym}C_{\mu\nu\lambda}&=C_{\nu\mu\lambda}=C_{\mu\lambda\nu},
\end{align}
where $R_{\mu\nu\lambda\sigma}$ -Riemann curvature, $C_{\mu\nu\lambda}$ is structure constants with index lowered by  $g_{\alpha \beta}$.

Using computer we found expression for the metric up to the 4th order in $t$. From equation~(\ref{curve}) we found expressions for~ $h_\mu(t)$ which still contain two parameters. After these equations (\ref{sym}) automatically satisfied. The solution of~(\ref{der}) fixes the value of one of the two parameters leaving only one. The fact that solution of equations~(\ref{curve}-\ref{sym}) have one parameter is in perfect agreement with~\cite{M.Saito1}.

Finally, one can find flat coordinates from the equation
\begin{equation}
\frac{\partial^2 s_\mu}{\partial t_\alpha \partial t_\beta}=\Gamma^\gamma_{\alpha\beta}\frac{\partial s_\mu}{\partial t_\gamma}.
\end{equation}
Since the metric found from equations (\ref{curve}-\ref{sym}) contains one parameter the flat coordinates will also contain a parameter. These results  are in perfect agreement up to the fourth order with the one of section \ref{sec conj} if we impose the constraint  $r_{1,14}=r_{2,15}$ .

\section{Resonances and modules of primitive form}
In this section we want to connect our results with one of \cite {M.Saito1}. At first we need to introduce some notation (in this section we will change several conventions in order for the formulas to coincide with \cite{LLS}). We will label weights of basis diagrams as $\sigma_i= {\deg e_i}/{\deg W_0}$. Basis in the ring must be ordered in such a way that $\sigma_1\le \sigma_2\le\dots\le \sigma_M$. It was proved in \cite {M.Saito1} that dimension $D$ of the moduli space of the primitive form (or number of free parameters) is given by the formula
\begin{equation}\label{M.Saito1}
D=\#\left\{(i,j)\,|\,p(i,j)\in \mathbb Z_{>0}, i+j<M+1\right\}+\#\left\{(i,j)\,|\,p(i,j)\in \mathbb Z_{>0}^{\text{odd}}, i+j=M+1\right\},
\end{equation}
where $p(i,j)=\sigma_i-\sigma_j$, $M$ is a dimension of the chiral ring, $\mathbb Z_{>0}$ are positive integers and $Z_{>0}^{\text{odd}}$ are odd positive integers.

Note that $p(i,j)$ being integer is exactly the condition of resonance ($[s_\mu]-[s_\nu]= 0 \mod d$ and $[s_\mu]\ne[s_\nu]$) we used in our paper. Consider a few simple examples.

\subsubsection*{Example 1. $\widehat {SU}(3)_3$.}
In this case dimension of the chiral ring is $M=10$ and there is one resonance for the $\sigma_1=\sigma_{\varnothing}=0$, $\sigma_{10}=\sigma_{\tableau{3 3}}=1$.
\begin{equation}
\quad p(10,1)=1, \quad 10+1= M+1=11 \rightarrow D=1.
\end{equation}
We see that in this case the number of the resonaces coincides with the number of the parameters
in the primitive form.

\subsubsection*{Example 2. $\widehat {SU}(3)_4$.}
In this case dimension of the chiral ring is $M=15$ and there are two resonances
\begin{align}
&(\sigma_1=0,\sigma_{14}=1)  &p(14,1)&=1,  & 14+1&=15<M+1=16,\\
&(\sigma_2=\frac 1 7,\sigma_{15}=\frac 8 7) & p(15,2)&=1,  & 15+2&=17>M+1=16.
\end{align}
One can see that in this case we have two resonances meanwhile  the primitive form possesses  only one parameter as demanded by (\ref{M.Saito1}).

\subsubsection*{Example 3. $\widehat {SU}(3)_5$.}

In this case the dimension of the chiral ring is $M=21$ and we have the resonances
\begin{align}
\varnothing \sim \tableau{5 3}, \tableau{4 4}, \quad  \quad  \quad
\tableau{1}  \sim \tableau{5 4}, \quad  \quad  \quad
\tableau{2},\tableau{1 1} \sim \tableau{5 5}.
\end{align}
The  computation of the moduli dimension gives
\begin{align}
&(\sigma_1=0,\sigma_{18}=1)  &p(18,1)&=1,  & 18+1&=19<M+1=22,\\
&(\sigma_1=0,\sigma_{19}=1)  &p(19,1)&=1,  & 19+1&=20<M+1=22,\\
&(\sigma_2=\frac 1 8,\sigma_{20}=\frac 9 8)  &p(20,2)&=1,  & 20+2&=22=M+1=22,\\
&(\sigma_3=\frac 2 8,\sigma_{21}=\frac {10} 8)  &p(21,3)&=1,  & 21+3&=24>M+1=22,\\
&(\sigma_4=\frac 2 8,\sigma_{21}=\frac {10} 8)  &p(21,4)&=1,  & 21+4&=25>M+1=22.
\end{align}
Thus the number of the resonances is five meanwhile  the dimension of the moduli space is three.

The extra resonances are the origin of extra parameters $r_{\mu,\nu}$ in the expressions for the flat coordinates obtained from (\ref {conj}).

\section{Conclusion}

The comparison of  two computations, performed in this work, shows that after imposing the constraint on the parameters  $r_{1,14}=r_{2,15}$  the both expressions for the flat coordinates coincide.

This coincidence  confirms the correctness of the Conjecture (\ref{conj}), now in the case when  there are one marginal and one irrelevant deformations in the addition to the relevant ones.

In the same time, this comparison leads to the interesting question about the nature of the constraints which have to be imposed on the  extra parameters $r_{\mu,\nu}$ which are predicted in \cite{M.Saito1}.

Recently a new perturbative method to compute the flat coordinates and the primitive form
has been suggested  in the papers \cite{LLS,LLSS}. It would be interesting to understand  the connection between this method and our approach.

Probably using the results of \cite{LLS,LLSS} can help to prove the Conjecture  (\ref{conj}) about the representation for flat coordinates through the oscillating integrals.

\vspace{7mm}

\noindent \textbf{Acknowledgements.} We thank A.~Givental for the useful discussions and  very valuable explanations on  the homology $ H_n({\mathbb C}^n, \operatorname{Re}W_{0}=-\infty)$.

The work was performed with the financial support of the Russian Science Foundation (Grant No.14-12-01383).

\section*{Appendix}
\subsection*{Expressions for $s_\mu$ up to second order in $t$}
Here we present expressions for the flat coordinates obeyed via the conjecture. In order for this answers to coincide with the direct computation of section \ref{dir comp} one must set $r_{1,14}=r_{2,15}$.
\begin{align*}
s_{15}&=t_{15}-(r_{1,14}+r_{2,15})t_{14} t_{15},\\
s_{14}&=t_{14}-r_{1,14}t_{14}^2 -r_{2,15}t_{13} t_{15} -r_{2,15}t_{12} t_{15} ,\\
s_{13}&=t_{13}+(3-r_{1,14})t_{13} t_{14} +(3-r_{2,15})t_{11} t_{15} -r_{2,15}t_{10} t_{15}, \\
s_{12}&=t_{12}-r_{1,14}t_{12} t_{14} -r_{2,15}t_{10} t_{15}, \\
s_{11}&=t_{11}+t_{13}^2+2 t_{12} t_{13}+ (2-r_{1,14})t_{11} t_{14}+2 t_{10} t_{14}-r_{2,15}t_9 t_{15} +(2-r_{2,15})t_8 t_{15}, \\
s_{10}&=t_{10}+\frac{3 t_{13}^2}{2}+3 t_{11} t_{14}-r_{1,14}t_{10} t_{14} +3 t_9 t_{15}-r_{2,15}t_8 t_{15} -r_{2,15}t_7 t_{15}, \\
s_{9}&=t_9+t_{11} t_{13}+t_{10} t_{13}+t_{10} t_{12}-r_{1,14}t_9 t_{14} +t_8 t_{14}+t_7 t_{14}-r_{2,15}t_6 t_{15} +t_5 t_{15},\\
s_{8}&=t_8+2 t_{11} t_{13}+2 t_{11} t_{12}+2 t_{10} t_{13}+2 t_9 t_{14}+(2-r_{1,14})t_8 t_{14} +(2-r_{2,15})t_6 t_{15} -r_{2,15}t_5 t_{15} , \\
s_{7}&=t_7+3 t_{11} t_{13}+3 t_9 t_{14}-r_{1,14}t_7 t_{14} -r_{2,15}t_5 t_{15} ,\\
s_{6}&=t_6+\frac{t_{11}^2}{2}+t_{10} t_{11}+\frac{t_{10}^2}{2}+t_9 t_{13}+t_8 t_{13}+t_8 t_{12}+t_7 t_{13}+(1-r_{1,14})t_6 t_{14} +t_5 t_{14}+(1-r_{2,15})t_4 t_{15}, \\
s_{5}&=t_5+t_{11}^2+2 t_{10} t_{11}+2 t_9 t_{13}+2 t_9 t_{12}+2 t_8 t_{13}+2 t_6 t_{14}-r_{1,14}t_5 t_{14}-r_{2,15}t_4 t_{15} -r_{2,15}t_3 t_{15}, \\
s_{4}&=t_4+t_9 t_{11}+t_9 t_{10}+t_8 t_{11}+t_8 t_{10}+t_7 t_{11}+t_6 t_{13}+t_6 t_{12}+t_5 t_{13}+(1-r_{1,14})t_4 t_{14} -r_{2,15}t_2 t_{15}, \\
s_{3}&=t_3-t_9 t_{11}+2 t_9 t_{10}+2 t_8 t_{11}+2 t_6 t_{13}-r_{1,14}t_3 t_{14} -r_{2,15}t_2 t_{15}, \\
s_{2}&=t_2+\left(\frac{1}{2} r_{2,15}-1\right)t_9^2 +t_8 t_9+\left(\frac{1}{2} r_{2,15}+\frac{1}{2}\right)t_8^2 +t_7 t_9+\frac{1}{2}r_{2,15} t_7^2 +(r_{2,15}+1)t_6 t_{11} +t_6 t_{10}+t_5 t_{11}\\&+ r_{2,15}t_5 t_{10}+ (r_{2,15}+1)t_4 t_{13}+r_{2,15}t_3 t_{12} +(r_{2,15}-r_{1,14})t_2 t_{14} ,\\
s_{1}&=t_1+(r_{1,14}-1)t_6 t_9 + (r_{1,14}+1)t_6 t_8+t_5 t_9+r_{1,14} t_5 t_8 r_{1,14}+r_{1,14}t_5 t_7 +(r_{1,14}+1)t_4 t_{11} +r_{1,14}t_4 t_{10} \\&+r_{1,14}t_3 t_{10} +r_{1,14}t_2 t_{13} + r_{1,14}t_2 t_{12}.
\end{align*}
\subsection*{Exact expressions for $t_{14}=t_{15}=0$}
Below we give expressions for flat coordinates with marginal and irrelevant deformations set to zero.
\begin{align*}
s_{15}&=t_{15}, \quad \quad \quad \quad \quad \,\,\,\,
s_{14}=t_{14}, \quad \quad\quad
s_{13}=t_{13},\quad \quad \quad
s_{12}=t_{12},\quad\quad
s_{11}=t_{11}+t_{13}^2+2 t_{12} t_{13},\\
s_{10}&=t_{10}+\frac3 2t_{13}^2, \quad \quad\quad
s_{9}=t_9+t_{11} t_{13}+t_{10} t_{13}+t_{10} t_{12}+2 t_{13}^3+3 t_{12} t_{13}^2,\\
s_{8}&=t_8+2 t_{11} t_{13}+2 t_{11} t_{12}+2 t_{10} t_{13}+5 t_{13}^3+5 t_{12} t_{13}^2+5 t_{12}^2 t_{13},\\
s_{7}&=t_7+3 t_{11} t_{13}+2 t_{13}^3+6 t_{12} t_{13}^2+2 t_{12}^3,\\
s_{6}&=t_6+\frac{t_{11}^2}{2}+t_{10} t_{11}+\frac{t_{10}^2}{2}+t_9 t_{13}+t_8 t_{13}+t_8 t_{12}+t_7 t_{13}+\frac{15}{2} t_{11} t_{13}^2+5 t_{11} t_{12} t_{13}+\frac{5}{2} t_{11} t_{12}^2+5 t_{10} t_{13}^2\\&+5 t_{10} t_{12} t_{13}+10 t_{13}^4+20 t_{12} t_{13}^3+10 t_{12}^2 t_{13}^2+\frac{20}{3} t_{12}^3 t_{13},\\
\end{align*}
\begin{align*}
s_{5}&=t_5+t_{11}^2+2 t_{10} t_{11}+2 t_9 t_{13}+2 t_9 t_{12}+2 t_8 t_{13}+8 t_{11} t_{13}^2+8 t_{11} t_{12} t_{13}+8 t_{10} t_{13}^2+8 t_{10} t_{12} t_{13}+4 t_{10} t_{12}^2\\&+\frac{43}3 t_{13}^4+24 t_{12} t_{13}^3+18 t_{12}^2 t_{13}^2,\\
s_{4}&=t_4+t_9 t_{11}+t_9 t_{10}+t_8 t_{11}+t_8 t_{10}+t_7 t_{11}+t_6 t_{13}+t_6 t_{12}+t_5 t_{13}+6 t_{11}^2 t_{13}+2 t_{11}^2 t_{12}+8 t_{10} t_{11} t_{13}\\&+4 t_{10} t_{11} t_{12}+4 t_{10}^2 t_{13}+2 t_{10}^2 t_{12}+4 t_9 t_{13}^2+4 t_9 t_{12} t_{13}+6 t_8 t_{13}^2+4 t_8 t_{12} t_{13}+2 t_8 t_{12}^2+2 t_7 t_{13}^2+4 t_7 t_{12} t_{13}\\&+30 t_{11} t_{13}^3+48 t_{11} t_{12} t_{13}^2+16 t_{11} t_{12}^2 t_{13}+\frac{16}{3} t_{11} t_{12}^3+\frac{80}{3} t_{10} t_{13}^3+32 t_{10} t_{12} t_{13}^2+16 t_{10} t_{12}^2 t_{13}+40 t_{13}^5+88 t_{12} t_{13}^4\\&+88 t_{12}^2 t_{13}^3+\frac{88}{3} t_{12}^3 t_{13}^2+\frac{44}{3} t_{12}^4 t_{13},\\
s_{3}&=t_3-t_9 t_{11}+2 t_9 t_{10}+2 t_8 t_{11}+2 t_6 t_{13}+3 t_{11}^2 t_{13}+3 t_{11}^2 t_{12}+6 t_{10} t_{11} t_{13}+3 t_{10}^2 t_{13}+3 t_{10}^2 t_{12}+3 t_9 t_{13}^2+3 t_8 t_{13}^2\\&+6 t_8 t_{12} t_{13}+3 t_7 t_{13}^2+3 t_7 t_{12}^2+24 t_{11} t_{13}^3+24 t_{11} t_{12} t_{13}^2+27 t_{11} t_{12}^2 t_{13}+17 t_{10} t_{13}^3+27 t_{10} t_{12} t_{13}^2+27 t_{13}^5\\&+\frac{135}{2} t_{12} t_{13}^4+45 t_{12}^2 t_{13}^3+45 t_{12}^3 t_{13}^2+\frac{9 t_{12}^5}{2}.
\end{align*}
Expressions for $s_1$ and $s_2$ are too lengthy and we don't present them here.
\subsection*{Primitive form up to second order in $t$}
We present the expression for the primitive form up to the second order in overall $t$. Note that decomposition (\ref{pr form}) is overdetermined since any polynomial of $e_\alpha$ can be reexpressed as polynomial of only $e_2$ and $e_3$. We used this freedom to express the primitive form linearly in $e_\alpha$.

\begin{align*}
\Omega&=\Big[1-r_{1,14} t_{14} +  \left(r_{1,14}^2-1\right)t_{14}^2+r_{1,14} r_{2,15} t_{13} t_{15} + \left(r_{1,14} r_{2,15}-2\right)t_{12} t_{15} \Big]e_1\\&+\Big[-r_{2,15}+\left(r_{2,15}^2+r_{1,14} r_{2,15}-3\right)t_{14}\Big]t_{15}e_2-3t_{15}^2e_3.
\end{align*}

\end{document}